\documentclass[letter,english]{article}
\usepackage{graphics,emulateapj,onecolfloat}

\begin{document}
\twocolumn
[
\title{The effect of bandpass uncertainties on component separation}
\author{Sarah Church${}^1$, Lloyd Knox${}^2$ and Martin White${}^3$}
\affil{${}^1$Department of Physics, Stanford University}
\affil{${}^2$Department of Physics, University of California,
Davis, CA 95616}
\affil{${}^3$Departments of Astronomy and Physics, University of California,
Berkeley, CA 94720}

\begin{abstract}
\noindent
\rightskip=0pt
Multi-frequency measurements of the microwave sky can be decomposed into maps
of distinct physical components such as the cosmic microwave background (CMB)
and the Sunyaev--Zel'dovich (SZ) effect.
Each of the multi--frequency measurements is a convolution of the spectrum on
the sky with the bandpass of the instrument.  Here we analytically calculate
the contamination of the component maps that can result from errors in our
knowledge of the bandpass shape.  We find, for example, that for {\sl Planck\/}
an unknown 10\% ramp across each band results in a CMB map
$\delta T = \delta T_{CMB} - 4.3 \times 10^{-3}\delta T_{SZ}$
plus the usual statistical noise.
The variance of this contaminant is more than a factor of 100 below the noise
variance at all angular scales and even further below the CMB signal variance.
This contamination might lead to an error in the velocity of rich clusters
inferred from the kinetic SZ effect, however the error is negligible,
${\cal O}(50{\rm km}\,{\rm s}^{-1})$, if the bandpass is known to 10\%.
Bandpass errors might be important for future missions measuring the CMB-SZ
correlation.
\end{abstract}
]

\section{Introduction} \label{sec:intro}

Small scale anisotropies in the cosmic microwave background can arise from
a number of sources.  In addition to the `primary' anisotropies, generated
at the surface of last scattering, secondary anisotropies and foregrounds
can contribute to the observed brightness fluctuations.  Separating the
components from multi-frequency observations is an important part of the
data reduction and interpretation.
Here we consider the effect of uncertainties in the frequency response of
the instrument within the observational bands and how that impacts our ability
to perform component separation.

\section{Model of the sky}

We assume for simplicity that the intensity in any given direction of the
sky is the sum of 5 components.  The first is the CMB itself with
specific intensity
\begin{equation}
  I_\nu = {dB_\nu\over d\nu} \propto {x^4 e^x\over (e^x-1)^2}
\end{equation}
where $B_\nu$ is a blackbody spectrum and $x=h\nu/k_BT_{CMB}\simeq\nu/56.84$GHz
is the dimensionless frequency.
The kinetic Sunyaev-Zel'dovich effect
(SZ; Sunyaev \& Zel'dovich \cite{SZ72,SZ80}; for recent reviews see
Birkinshaw \cite{Bir} and Rephaeli \cite{Rep}), arising from the motion of
ionized gas with respect to the rest-frame of the CMB, has the same frequency
dependence as the CMB signal.
The second component is the thermal SZ effect -- one of the primary sources of
secondary anisotropies in the CMB on small angular scales.
Ignoring relativistic corrections, the change in the (thermodynamic)
temperature of the CMB resulting from scattering off non-relativistic
electrons is
\begin{eqnarray}
{\Delta T\over T} &=&
  \phantom{-2}y \left( x{{\rm e}^x+1\over {\rm e}^x-1}-4 \right) \\
  &\simeq& -2y\qquad \mbox{for }\ x\ll 1\, ,
\end{eqnarray}
where the second expression is valid in the Rayleigh-Jeans limit and $y$
is the Comptonization parameter which is proportional to the integrated
electron pressure along the line of sight.

\begin{table}
\begin{center}
\begin{tabular}{ccccc}
Number & Frequency & $\Delta\nu/\nu$ & Beam & Noise \\ \hline
  1    &    30     &    0.2          & 33   & 5.5   \\
  2    &    44     &    0.2          & 24   & 7.4   \\
  3    &    70     &    0.2          & 14   & 13    \\
  4    &   100     &    0.2          & 10   & 21    \\
  5    &   100     &    0.3          & 9.2  & 5.5   \\
  6    &   143     &    0.3          & 7.1  & 6     \\
  7    &   217     &    0.3          & 5.0  & 13    \\
  8    &   353     &    0.3          & 5.0  & 40    \\
  9    &   545     &    0.3          & 5.0  & 400
\end{tabular}
\end{center}
\caption{The parameters assumed for {\sl Planck}.  The central frequency
is quoted in GHz, the beam size in arcminutes and the noise (thermodynamic
temperature fluctuation in a square pixel of side `Beam') in $\mu$K assuming
a 12 month integration.  We do not use the 850GHz channel of {\sl Planck\/}
here since our focus is on components at lower frequency.}
\label{tab:param}
\end{table}

We also include dust, bremsstrahlung (or free-free) emission and synchrotron
radiation following Knox (\cite{knox99}, see below).  Specifically for
the dust we assume the spectral dependence of a modified blackbody with
an emissivity index of 2 and a temperature of $18$K.
For the bremsstrahlung and synchrotron we assume power-laws in frequency
with indices $-0.16$ and $-0.8$ respectively (Bennett et al.~\cite{Ben}).
Our model does not include a contribution from extragalactic point sources.
For the purposes of this analysis we hold the spectral indices of the
components fixed (and known).  Though we do not expect this to be true of
real astrophysical foregrounds, the impact of this on our analysis is
negligible.

\begin{figure}
\begin{center}
\resizebox{3.5in}{!}{\includegraphics{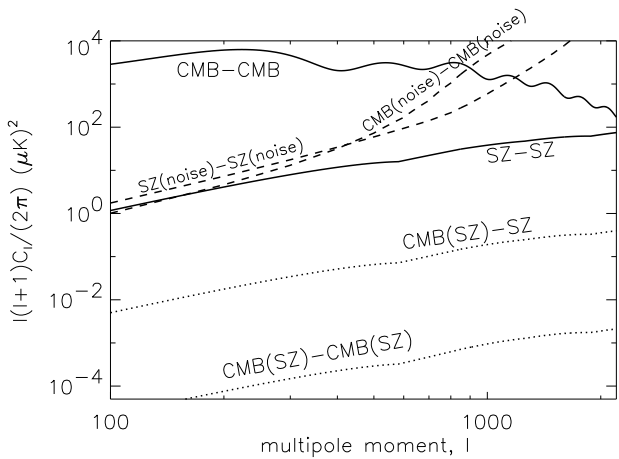}}
\end{center}
\caption{\footnotesize%
The angular power spectra, $\ell(\ell+1)C_\ell/(2\pi)$, of the CMB and
thermal SZ signals (solid) fluctuations and the  CMB and thermal SZ map
noise (dotted). 
The tSZ contamination of the CMB map, CMB(SZ), correlated with itself
(with SZ) are shown as the dotted curves. We assumed a 10\%
ramp error on all {\sl Planck\/} bandpasses.}
\label{fig:cl}
\end{figure}

\section{Method} \label{sec:method}

\begin{table*}
\begin{center}
\begin{tabular}{c|rrrrr}
     &  CMB  &    tSZ   &   Dust   &   Sync   &  Brem    \\ \hline
CMB  &  0.0  &   $-4.3\times 10^{-3}$  &  $-2.5\times 10^{-1}$ &   $ 8.3\times 10^{-5}$  &  $ 1.4\times 10^{-5}$\\
tSZ  &  0.0  & $3.7\times 10^{-3}$  &  $ 6.4\times 10^{-1}$  &  $-2.2\times 10^{-4}$   &$ -8.6\times 10^{-5}$\\
Dust &  0.0  & $7.2\times 10^{-7}$  &   $2.1\times 10^{-2} $  & $ 8.6\times 10^{-7}$    & $4.7\times 10^{-8}$\\
Sync &  0.0  & $6.9\times 10^{-3}$   & -4.3 & $ -2.6\times 10^{-3}$  &  $ 4.1\times 10^{-4}$\\
Brem &  0.0  & $-6.3\times 10^{-3}$   & $ 4.0$  &  $-8.1\times 10^{-4} $ &  $-4.9\times 10^{-3}$
\end{tabular}
\end{center}
\caption{\footnotesize%
The component-mixing matrix, $M_{\alpha\beta}$, for a 5 component model
(cmb, sz, dust, synchroton, bremsstrahlung) if all of the frequency channels
have an undetected $+10\%$ ramp (see text).}
\label{tab:rampall1}
\end{table*}

To understand the effect of an uncertainty in the response of the instrument
to a particular signal let us postulate the following situation.
We imagine that the sky is observed at $N$ frequencies, with measurements
$\theta_i$ ($i=1,\cdots,N$).  The signal is the sum of $M$ components
with amplitudes $s_\alpha$ ($\alpha=1,\cdots,M$) such that
\begin{equation}
  \theta_i = \sum_\alpha f_{i\alpha} s_\alpha + n_i
\label{eqn:model}
\end{equation}
where $n_i$ is the noise in channel $i$.  We write our linear estimator
of $s_\alpha$ as
\begin{equation}
  \widehat{s}_\alpha = \sum_i W_{i\alpha} \theta_i \qquad .
\end{equation}
Requiring $\widehat{s}$ to be an unbiased estimator and minimizing the
rms residual, $\langle (\widehat{s}-s)^2\rangle$, gives
\begin{equation}
  W_{i\alpha} = \sum_{\beta j}
    F^{-1}_{\alpha\beta}
    f_{j\beta} N^{-1}_{ji}
\end{equation}
where $N_{ij}\equiv\langle n_i n_j\rangle$ and
$F_{\alpha \beta} = \sum_{ij} f_{i\alpha}  N^{-1}_{ij}f_{j\beta}$
is the Fisher matrix for the $s_\alpha$.
In the case of diagonal noise, $N_{ij} = \sigma_i^2\delta_{ij}$, the
Fisher matrix simplifies to
\begin{equation}
  F_{\alpha \beta} = \sum_i f_{i\alpha}f_{i\beta}/\sigma_i^2
\end{equation}
and the estimator can be written as
\begin{equation}
  \widehat{s}_\alpha = \sum_{i\beta} F^{-1}_{\alpha \beta}
    \ {f_{i\beta} \over \sigma_i^2} \theta_i  .
\label{eqn:umv}
\end{equation}
The covariance matrix of the statistical errors for this estimator is
$\langle(\hat s_\alpha-s_\alpha)(\hat s_\beta-s_\beta)\rangle =
F^{-1}_{\alpha \beta}$.

\begin{table*}
\begin{center}
\begin{tabular}{c|rrrrr}
     &  CMB  &    tSZ   &   Dust   &   Sync   &  Brem    \\ \hline
CMB  &  0.0  & -0.00032 & -0.00005 &  0.00001 &  0.00000 \\
tSZ  &  0.0  &  0.00355 &  0.00644 & -0.00019 & -0.00019 \\
Dust &  0.0  &  0.00002 &  0.02191 &  0.00010 &  0.00010 \\
Sync &  0.0  &  0.01387 & -0.09358 & -0.00373 &  0.00373 \\
Brem &  0.0  & -0.00747 &  0.05165 & -0.00025 & -0.00025
\end{tabular}
\end{center}
\caption{\footnotesize%
The normalized mixing matrix, $W_{\alpha\beta}$, for a 5 component model
(cmb, sz, dust, synchroton, bremsstrahlung) if all of the frequency channels
have an undetected $+10\%$ ramp (see text).}
\label{tab:rampall2}
\end{table*}

We are interested in the effects on our component separation of a band error
which causes the actual bandpass to deviate from the design bandpass by
$\delta f$.
Using Eq.~(\ref{eqn:umv}) with the design bandpass, but replacing
$\theta_i$ with the right-hand side of Eq.~(\ref{eqn:model}) evaluated with
the actual bandpass, and subtracting off the true signal we find
\begin{equation}
  \delta s_\alpha \equiv
   \langle \widehat{s}_{\alpha}-s_\alpha \rangle = 
   \sum_\beta M_{\alpha \beta} s_\beta
\end{equation}
where $M$ is the `component mixing matrix' given by
\begin{equation}
  M_{\alpha \beta} \equiv \sum_{i\gamma}
     F_{\alpha \gamma}^{-1}\ {f_{i\gamma}\delta f_{i\beta}\over\sigma_i^2}
  \qquad 
\end{equation}
and $\delta f_{i\beta}$ is the difference between the design and actual
bandpasses.  The bandpass uncertainty then induces a relative rms error on
component $\alpha$ from component $\beta$ characterized by
\begin{equation}
  W_{\alpha\beta} \equiv {M_{\alpha\beta}
  \langle s_\beta^2 \rangle^{1/2}\over \langle s_\alpha^2\rangle^{1/2}}
  \qquad .
\end{equation}

How is $\delta f_{i \alpha}$ related to the bandpass error?  If component
$\alpha$ has frequency dependence $g_\alpha(\nu)$ then
\begin{equation}
  f_{i\alpha} = \int d\nu\ g_\alpha(\nu)
  \left[\vphantom{\int} h_i(\nu)+\delta h_i(\nu)\right]
\end{equation}
where $h_i(\nu)+\delta h_i(\nu)$ is the total bandpass, with the latter term
the error.  It is easy to see that an error in the amplitude of the bandpass,
$\delta h_i(\nu)\propto h_i(\nu)$, will have no effect on component separation
since it will `calibrate out'.
In general we shall model this calibration process by demanding that
\begin{equation}
  f_{i0} = \int d\nu\ g_0(\nu)
  \left[\vphantom{\int} h_i(\nu)+\delta h_i(\nu)\right] = 1
  \quad \forall i
\end{equation}
where component $0$ is the CMB. 
We normalize $g_\alpha(\nu=30~\rm{GHz})=1$ for all components $\alpha$.
Note that this means $s_\alpha$ is the amplitude of component $\alpha$
at 30 GHz.

Note that the above treatment applies in both pixel space and spherical
harmonic space.  In pixel space the $\sigma_i$ should all be calculated for
the same pixel size.
In spherical harmonic space, the $\sigma_i$ (interpreted as errors on the
beam--deconvolved maps) are $\ell$--dependent:
\begin{equation}
  \sigma_i(\ell)= \hat{\sigma}_i {\vartheta}_i
    \exp\left[{1\over 2}\left({\ell\vartheta_i\over 2.355}\right)^2\right]
\end{equation}
where ${\vartheta}$ and $\hat{\sigma}$ are the beam and noise defined in
Table~\ref{tab:param} with ${\vartheta}$ converted to radians.

\section{Results}

As an example let us consider simple shifts in the bandpass, which we shall
model as a top-hat of width $\Delta\nu$ about the central frequency as given
in Table~\ref{tab:param}.  To gain intuition let us restrict ourselves to a
$2\times 2$ subspace consisting only of CMB and tSZ signals with a $1$GHz
shift in the $217$GHz channel to higher frequency.  In this case (at angular
scales larger than the largest beam)
\begin{equation}
  M_{\alpha\beta} = \left( \begin{array}{cc}
                     0.0 &           -0.00391 \\
                     0.0 & \phantom{-}0.00522
                     \end{array}\right)
\end{equation}
which indicates that the CMB channel does not `contaminate' the any of the
signals but there is leakage from the tSZ channel.  That $M_{\alpha 0}=0$
is a direct result of our calibration procedure which enforces
$\delta f_{i 0}=0$.  Note that for other calibration sources typically
used by ground--based experiments, e.g. planets, the CMB {\em can}, via 
bandpass errors, contaminate other components.

A more realistic scenario is that the bandpass frequency be quite well
determined, but the amplitude of the response as a function of frequency
be somewhat uncertain.  This holds for both HEMTs (M.~Seiffert, private
communication) and Bolometers (P.~Ade, private communication).
As a simple model of this effect we introduce a linear `ramp' into our
otherwise top-hat bandpasses with a change in amplitude of 10\% across the
band.  The results of including a $+10\%$ ramp in all of the channels with
our 5 component model is given in Table~\ref{tab:rampall1}.

The elements of the mixing matrix in Table~\ref{tab:rampall1} tell us,
for example, that our CMB map, $\hat s_0$ will have a contribution from
dust of $-0.25 s_2$, where $s_2$ is the dust amplitude at 30 GHz.
Fortunately the dust amplitude is very low at 30 GHz!

The importance of this contamination is easier to read from the normalized
mixing matrix, $W_{\alpha \beta}$.  To calculate $W_{\alpha \beta}$ we need
to know the rms sky fluctuations of the various components.
We use the model described in Knox~(\cite{knox99}) and references therein,
most notably Bouchet \& Gispert~(\cite{bouchet99}).
The exception is our SZ power spectrum which we take from
White, Hernquist \& Springel (\cite{WHS}) and extend to $\ell<400$ by assuming
$C_\ell = C_{400}$ appropriate for the Poisson dominated regime
(see Fig.~\ref{fig:cl}).

In Table~\ref{tab:rampall2} we assume $s_\alpha$ are the amplitudes of an
$\ell=500$ spherical harmonic, so that
$\langle s_\alpha^2 \rangle = C_\ell^\alpha$.
We find that all the elements are quite small; the largest contribution
to the CMB map comes from SZ and the rms of this contaminant is .05\% of
the CMB signal rms.
The galactic contaminants of the CMB map, $W_{0\beta}$ for $\beta=2,3,4$
greatly increase with increasing angular scale.
Even so, they are very small at all angular scales; at $\ell=2$
$W_{0\beta}$ = -0.006, 0.0002 and $2\times 10^{-5}$ for dust, synchrotron and
bremstrahlung respectively.

In Fig.~\ref{fig:cl} we show how the SZ contamination of the CMB map affects
the CMB power spectrum and the CMB-SZ cross--correlation power spectrum.
Denoting the CMB contaminant from SZ as $a_{lm}^{\rm CMB(SZ)}$ we have
\begin{equation}
  C_\ell^{\rm CMB(SZ)-SZ} \equiv
  \langle a_{\ell m}^{\rm CMB(SZ)}(a_{\ell m}^{\rm SZ})^*\rangle
  = M_{01}(\ell) C_\ell^{\rm SZ-SZ}.
\end{equation}
The $\ell$-dependence of the mixing matrix arises from the $\ell$-dependence
of the noise in the beam--deconvolved maps but is quite mild:  $M(0,1)$
monotonically decreases from $-4 \times 10^{-3}$ to its high $\ell$ asymptote
of $-5.3 \times 10^{-3}$.
In general the cross--correlations are the most affected since they are first
order in the component mixing matrix.
Fortunately, for the 10\% ramp error, the bandpass error--induced
cross--correlation is well below the level of the {\sl Planck\/} noise.  

Our results are specific to the conservative component separation procedure
we have assumed.  Other methods can reduce the statistical noise by including
assumptions about the statistical properties of the various components
(Tegmark \& Efstathiou \cite{tegmark96};
 Bouchet, Gispert \& Puget \cite{bouchet95};
 Hobson et al.~\cite{hobson98}).
The results will also differ in detail if one adopts a more realistic (and
more complicated) model of the foregrounds.  However, we do not expect the
component mixing matrix to be qualitatively different for these different
procedures or for more realistic foreground modeling.
The fact that the two--component model and the five--component model give
similar results for both $M(0,1)$ and $M(1,1)$ we take as evidence of this
robustness.  

The bandpass uncertainties can lead to systematic errors in the velocities of
clusters inferred from the kinetic SZ effect.
The mixing between thermal SZ and primary CMB on a cluster of optical depth
$\tau$ and temperature $T_e$ induces an error
\begin{equation}
  \delta T \simeq -8\times 10^{-3}\ \tau\left({kT_e\over m_ec^2}\right)
\end{equation}
in the CMB signal.
If the contamination were all erroneously attributed to kinetic SZ from the
moving cluster, it would bias the inferred velocity by
$v=8\times 10^{-3}(kT_e/m_ec^2)c\approx 50{\rm km}\,{\rm s}^{-1}$
for a rich cluster.  This bias is negligible compared to other sources of
uncertainty for individual cluster velocities 
(\cite{haehnelt96,nagai02,holder02})
but is comparable to errors that might be achievable by {\sl Planck\/} on
bulk flows in $10^6 h^{-3}$ Mpc$^3$ volumes (\cite{aghanim01}).
This systematic contaminant of the bulk flows will appear as a $T_e$--weighted
monopole.  Such a pattern is not expected cosmologically, would be evidence of
bandpass errors and could be removed from the data with negligible residuals.
Note that while we have ignored relativistic corrections to the SZ distortion
for the purposes of estimating the magnitude of the effect, they will be
important in the actual analysis of the data (\cite{diego02}).

\section{Conclusions} \label{sec:conclusions}

We can conclude that bandpass errors at the ${\cal O}(10\%)$ level are
acceptably small for {\sl Planck}.
More sensitive experiments though may have more stringent requirements
on the quality of the bandpass measurements because an overall scaling of
the sensitivity of each channel leaves the component mixing matrix unchanged.

\bigskip
\acknowledgments
We thank P.~Ade, G.~Holder, C.~Lawrence and M.~Seiffert for useful
conversations.
M.W. was supported by a NASA Astrophysical Theory Grant, the NSF and a
Sloan Fellowship.  L.K. was supported by NASA NAG5-11098.  We thank
the KITP (supported by NSF PHY99-07949) for their hospitality.

\end{document}